\documentclass[prb,aps,10pt,amsmath,amssymb,twocolumn,floatfix,superscriptaddress]{revtex4-2}
\usepackage{graphicx,bm,orcidlink}

\begin{document}
\title{Spontaneous Decay of High Energy Composite Fermions}
\author{Mytraya Gattu~\orcidlink{0000-0001-6994-389X}}
\thanks{Current address: Research Laboratory of Electronics, Massachusetts Institute of Technology, Cambridge, Massachusetts 02139, USA; email: \texttt{mytraya@mit.edu}.}
\affiliation{Department of Physics, 104 Davey Lab, Pennsylvania State University, University Park, Pennsylvania 16802, USA}
\author{J. K. Jain~\orcidlink{0000-0003-0082-5881}}
\affiliation{Department of Physics, 104 Davey Lab, Pennsylvania State University, University Park, Pennsylvania 16802, USA}
\affiliation{Center for Theory of Emergent Quantum Matter, Pennsylvania State University, University Park, Pennsylvania 16802, USA}
\affiliation{Lodha Theoretical Physics Institute, 17th Floor, Lodha NCP Supremus, Wadala (E), Mumbai 400037, India}
\date{\today}
\begin{abstract}
An electron, when added to a fractional quantum Hall fluid, famously splits into fractionally charged composite-fermion quasiparticles. We show here that a high energy composite fermion carrying a fractional charge decays into a collection of particles, including quasiparticles, quasiholes, and even whole electrons. 
\end{abstract}
\maketitle

{\bf Introduction:} A clear signature of the strongly correlated nature of the fractional quantum Hall effect (FQHE) is that its ``fundamental particles" are composite fermions (CFs)~\cite{Jain89,Jain07,Jain20}, topologically distinct from electrons. Much of the vast phenomenology of the FQHE is straightforwardly understandable in terms of noninteracting CFs: the observed sequences $\nu=n/(2pn\pm 1)$ as integer quantum Hall effect (IQHE) of CFs~\cite{Jain89} and the non-FQH state at half-filling as a Fermi liquid metal of CFs~\cite{Halperin93,Halperin20,Shayegan20}.  In particular, when an electron is added to a FQHE state, it splits into many fractionally charged CFs. 

The CF paradigm has also suggested explicit wave functions. The construction proceeds through the following steps, taking the example of the incompressible ground state at $\nu=n/(2pn+1)$: 
\begin{enumerate}
    \item begin with the known IQHE state $\Phi_n$ at filling factor $n$;
    \item attach vortices to each electron by multiplication by a Jastrow factor $\prod_{j<k}(z_j-z_k)^{2p}\sim \Phi_1^{2p}$ where $z_j=x_j-iy_j$ represents the position of the $j$th electron as a complex quantity; and 
    \item project the product into the LLL to obtain a wave function residing strictly in the LLL, as appropriate in the limit of high magnetic fields.
\end{enumerate}
This produces the Jain states $\Psi_{\nu=n/(2pn\pm 1)}={\cal P}_{\rm LLL} \Phi_n \Phi_1^{2p}$. (Multiplication by $\Phi_1^{2p}$ followed by LLL projection is referred to as composite-fermionization.)
Wave functions for charged and neutral excitations can be similarly constructed by beginning with the charged and neutral excitations of the IQHE state. These have been found to be exceedingly accurate in comparisons to exact eigenstates known for systems containing up to 16-18 particles~\cite{Jain07}.

{\bf A paradox:}
Our study was motivated by the following paradox. Two models have been proposed for the neutral excitations of an incompressible FQH state. The one by Girvin, MacDonald, and Platzman (GMP)~\cite{Girvin86,Girvin85} models the elementary neutral excitation mode as a projected density wave of {\it electrons}, given by  
\begin{equation} 
\Psi_{n/(2pn+1)}^{{\rm GMP},\bm{q}}\equiv {\cal P}_{\rm LLL}\;\left[\rho_{_{\bm{q}}}\; \Psi_{n/(2pn+1)}\right],
\end{equation} 
where $\rho_{_{\bm{q}}}$ is the electron density operator, $\Psi_{n/(2pn+1)}$ is the ground state at filling factor $\nu=n/(2pn+1)$, and ${\cal P}_{\rm LLL}$ denotes projection into the LLL. A knowledge of the ground state by other means is a prerequisite in this approach. Additional neutral excitations can be constructed by multiple applications of the density operator. 

The second model for neutral excitations relies crucially on the physics of the FQHE. It makes use of the fact that the incompressible ground states at $\nu=n/(2pn+1)$ are very accurately represented as the Jain states $\Psi_{n/(2pn+1)}={\cal P}_{\rm LLL} \Phi_n\Phi_1^{2p}$, where $\Phi_n$ is the Slater determinant wave function of $n$ filled LLs.  Neutral excitations are naturally modeled as CF excitons (CFEs), obtained by composite-fermionizing electron-hole excitons at $\nu=n$. A basis for single CFEs is given by the wave function 
\begin{equation} 
\{\Psi^{{\rm CFE},\alpha}_{n/(2pn+1)}\}=\{{\cal P}_{\rm LLL} \Phi^{{\rm e-h},\alpha}_n\; \Phi_1^{2p}\},
\end{equation}
where $\Phi^{{\rm exciton},\alpha}_n$ are all possible states at $\nu=n$ where an electron has been excited from an occupied LL to an unoccupied LL. This produces a complete basis for single CFEs. Past work has demonstrated that the actual low-energy neutral excitations, e.g. those seen in exact diagonalization studies, are very accurately represented as the lowest CFE, obtained by composite-fermionizing the lowest energy electron-hole exciton at $\nu=n$. The GMP mode has been found to reduce to the CFE in the long wave length limit. 

Ref.~\cite{Balram24} demonstrated exactly that the GMP mode defined with the Jain ground state resides within the space spanned by the single CF basis:
\begin{equation}\label{eq:theorem}
    \Psi_{n/(2pn+1)}^{{\rm GMP},\bm{q}} \in \{\Psi^{{\rm CFE},\alpha}_{n/(2pn+1)}\}\;.
\end{equation}
That is, the GMP mode can be expressed exactly as a linear combination of the single CFE modes. Ref.~\cite{Balram24} demonstrated this result numerically for all cases that are amenable to exact computation. We have recently proven Eq.~(\ref{eq:theorem}) analytically~\cite{Gattu-unp}. While pleasing, this result is rather surprising for the following reason. The GMP mode creates an {\it electron-hole} pair, while the CF exciton creates a CF particle-hole pair. An electron added to the $\nu=n/(2pn+1)$ state corresponds to $2pn\pm 1$ excited CFs, given that each excited CF has a charge $-e/(2pn\pm 1)$ associated with it. Removing an electron similarly creates $(2pn+1)$ CF holes. This counting would seem to imply that the GMP mode would belong in the space consisting of $2pn\pm 1$ CFEs. How do we reconcile this physics with the theorem expressed by Eq.~(\ref{eq:theorem})?

The problem can be expressed more physically by considering an extreme limit. At small wave vectors, $k\to 0$, where in the LLL the electron-hole separation $k\ell_B^2$ is small ($\ell_B=\sqrt{\hbar c/eB}$ is the magnetic length), the positive and negative charges of the electron and the hole are overlapping and therefore it may be possible to represent the electron-hole excitation as a linear superposition of single CF particle-hole excitations. However, at a large wave vector, the GMP mode represents a far separated electron hole pair. In fact, in the spherical geometry, as shown below, for certain wave vectors, the GMP mode creates an electron and a hole at antipodal points in the spherical geometry. An electron is a rather complex object from the perspective of composite fermions, as seen theoretically~\cite{Peterson05,Gattu23,Pu23A} as well as experimentally~\cite{Hu23}. The paradox may thus be stated as:

\underline{Paradox}: How can a pair of far separated electron and hole, each of which has a unit charge, be exactly represented in terms of a pair of far separated CF-particle and a CF-hole, each of which has a fractional charge?

{\bf The role of LLL projection:}
We see below that the resolution of the paradox is related to another aspect of the Jain wave functions, namely LLL projection. The unprojected Jain wave function $\Phi_n\Phi_1^2$ explicitly has good correlations built into it, as it vanishes with a high power of the distance between two particles as they are brought close to one another. It, however, is not restricted to the LLL.  The LLL restriction is by no means a necessary aspect of the FQHE physics, as FQHE is always seen in a finite $B$, which implies that the exact states always have a non-zero occupation of higher LLs.  Nonetheless, the LLL restriction is useful as it provides a nice theoretical limit where the dimension of the Hilbert space is finite for a finite system and exact eigenstates can be obtained on the computer. In fact, the LLL projected wave functions also provide a useful starting point for treating LL mixing~\cite{Zhang16,Zhao18,Gattu26}.

It is assumed that the act of LLL projection retains the correlations of the unprojected wave functions as best as possible, and also does not alter the topological properties. Originally, this was justified by the observation that the unprojected wave functions have only a small admixture with higher LLs~\cite{Jain89,Trivedi91,Kamilla97B}, which suggests that the LLL projection is only a gentle perturbation. An adiabatic scheme connecting the unprojected and projected Jain states has been implemented explicitly for the $2/5$ state for small systems~\cite{Rezayi91}. Most important proof-is-in-the-pudding support comes from the fact that the LLL projected wave functions are remarkably close to the exact Coulomb eigenstates, which establishes the validity of the LLL projected Jain states. As a result, even though there is no rigorous proof, it has been widely accepted that the LLL projection does not cause non-perturbative changes. 

The most natural method for LLL projection is to expand the product wave function in the Slater determinant basis, and then retain only those Slater determinants which reside fully in the LLL~\cite{Dev92a,Wu93}. This is referred to as the ``direct projection." A disadvantage of this projection method is that the basis size grows exponentially with the number of CFs, and soon outgrows the disk space on any computer; in practice, the direct projection can be performed only for systems with $\sim$10 or fewer particles~\cite{Dev92a,Wu93}. A more convenient method is what is known as the Jain-Kamilla (JK) projection~\cite{Jain97,Jain97b}, which has been found to produce wave functions for the ground and low-energy excited states that, while not identical, are extremely close to the direct-projected ones, and, with the recently developed quaternion representation of the wave functions, can be implemented for systems with as many as 1000 particles~\cite{Gattu25}. 

\begin{figure*}[!t]
  \centering
  {\large $\nu=1/3$}\par\vspace{2pt}
  \begin{tabular}{@{}l@{\hspace{\columnsep}}l@{}}
    \textbf{(a)} & \textbf{(b)} \\
    \includegraphics[width=246pt]{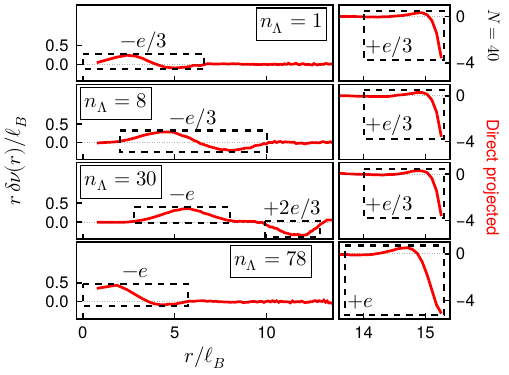} &
    \includegraphics[width=246pt]{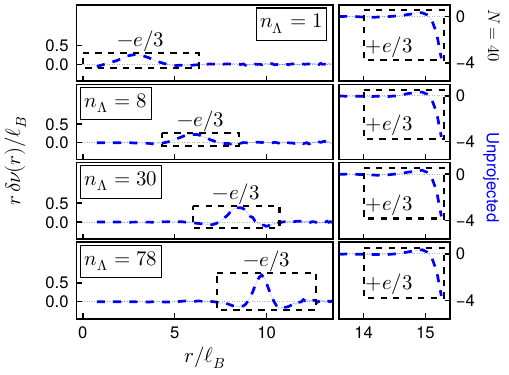} \\[4pt]
    \textbf{(c)} & \textbf{(d)} \\
    \includegraphics[width=246pt]{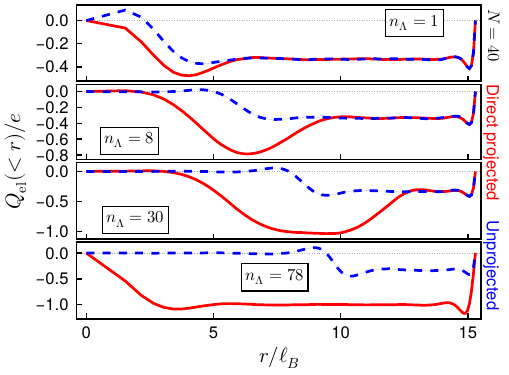} &
    \includegraphics[width=246pt]{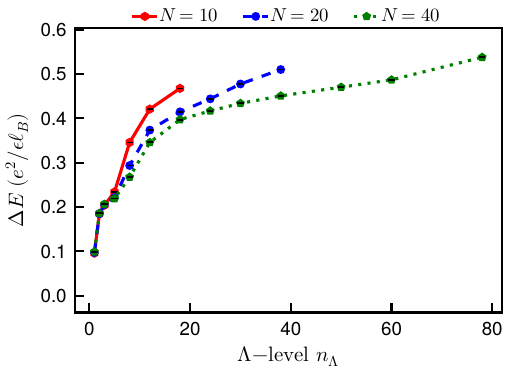}
  \end{tabular}
  \caption{
  CF exciton at $\nu=1/3$ in the spherical geometry, created by exciting a CF from the lowest $\Lambda$ level at the south pole to the $n_\Lambda\geq1$ $\Lambda$ level at the north pole (see text for details). At the maximum index, $n_\Lambda=2(N-1)$, the state coincides with the maximum-angular-momentum Girvin--MacDonald--Platzman mode.
  (a) Area-weighted excess density $r\delta\nu(r)/\ell_B$ above the uniform FQH ground state for the direct-projected state, for $N=40$ and $n_\Lambda=1,8,30,78$. (b) Same for the unprojected state.
  (c) Enclosed electric charge $Q_{\rm el}(<r)/e$ for the states in (a) (solid) and (b) (dashed). The charges of the projected CF particle--hole pair evolve from $\pm e/3$ to $\pm e$. In the intermediate regime, a localized charge $-e$ is accompanied by a compensating charge $+2e/3$ farther out and the original $+e/3$ quasihole at the south pole, preserving overall neutrality.
  (d) Excitation energy $\Delta E$ of the direct-projected state as a function of $n_\Lambda$ for $N=10$, 20, and 40, in units of $e^2/\epsilon\ell_B$, where $\epsilon$ is the dielectric constant and $\ell_B=\sqrt{\hbar c/eB}$ is the magnetic length. The energy rises with $n_\Lambda$; in the thermodynamic limit, the energy at the maximum index equals that of a far-separated electron--hole pair.
  In each case, the Monte Carlo error is smaller than the line width.}
  \label{fig:projection-density-charge}
  \label{fig:density-vs-lambda}
  \label{fig:projected-coulomb-energy}
\end{figure*}

{\bf Resolution:} We now return to the paradox. An important aspect of the theorem is that it is valid only for the direct-projected CFEs. Fortunately, the direct-projected CFEs of the $\nu=1/3$ state may be evaluated for large systems. All our results below are for the $\nu=1/3$ state. We will work in Haldane's spherical geometry~\cite{Haldane83} which considers $N$ electrons on the surface of a sphere subject to a radial magnetic field producing a net flux $2Q\phi_0$, with $\phi_0=hc/e$. The $\nu=1/3$ state occurs at $2Q=3N-3$. The GMP mode or the CF exciton mode is labeled by its angular momentum, which is related to the wave vector by $k=\sqrt{L(L+1)}/R$, where $R=\sqrt{Q}$ is the radius of the sphere in units of the magnetic length $\ell_B$, set to unity below. Each electron in the LLL has angular momentum $l=Q$, and the largest angular momentum for the GMP mode is $L=2Q$. For $L=2Q=L_z$ the GMP mode creates, relative to the uniform density ground state, a charge $-e$ at the North Pole and a charge $+e$ at the South Pole. 

In Fig.~\ref{fig:density-vs-lambda} we show results for the $\nu=1/3$ state of  $N=40$ particles at $Q=117/2$. The largest angular momentum for the GMP mode is $2Q=117$. The CF theory maps the state into $\nu^{\rm CF}=1$ of CFs at $Q^{\rm CF}=(N-1)/2=39/2$.  Here an exciton consists of a hole in the $0$th LL with angular momentum $Q^{\rm CF}$ and an electron in the $n_\Lambda$th excited LL with angular momentum $Q^{\rm CF}+n_\Lambda$. We consider the state with $L=L_z=2Q^{\rm CF}+n_\Lambda=39+n_\Lambda$. Composite-fermionizing it produces a CF exciton with a QH in the lowest $\Lambda$L and a QP in the $n_\Lambda$th $\Lambda$L.

Figs.~\ref{fig:density-vs-lambda} (a) and (b) show the density, integrated over the azimuthal angle, as a function of the distance from the North Pole. Figure~\ref{fig:density-vs-lambda}(c) shows the cumulative charge excess relative to the uniform density of the ground state, in the region within distance $r$ from the North Pole. 

\begin{figure*}[!t]
  \centering
  \includegraphics{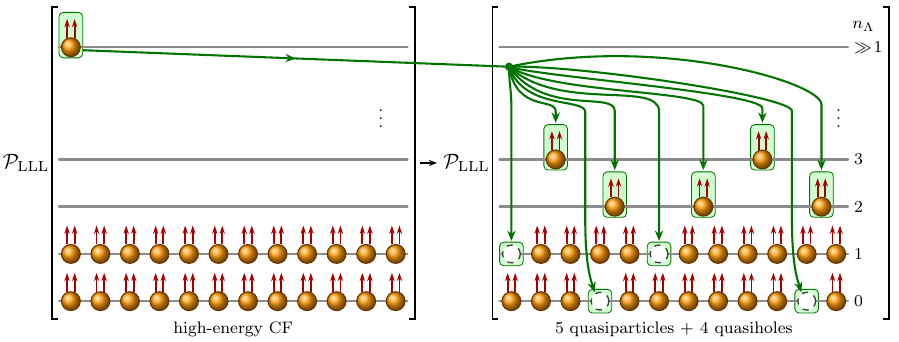}
  \caption{Schematic depiction of the decay of a high-energy composite fermion (CF) occupying a high $\Lambda$ level into many low-energy CF particles and holes, highlighted in green. The figure is drawn for filling factor $\nu=2/5$, for which each excited CF has charge $-e/5$ associated with it and a CF hole $+e/5$; the high-energy CF decays into five quasiparticles and four quasiholes.}
  \label{fig:cf-decay-schematic}
\end{figure*}

We find three kinds of behaviors as we increase $n_\Lambda$.
\begin{enumerate}
    \item The lowest energy CFE, which has a QP in the $n_\Lambda=1$ $\Lambda$L, is shown in the first enclosed-charge panel: it contains a charge $-e/3$ QP at the North Pole and a charge $e/3$ QH at the South Pole. This behavior persists for small $n_\Lambda$.
    \item  With increasing $n_\Lambda$, the charge excess around the North Pole becomes $-e$, followed by a region of charge $+2e/3$, which is  then followed by a charge $+e/3$ at the South Pole. Here the LLL projection of a QP and QH yields a whole electron, two charge $e/3$ QHs and then another charge $e/3$ QH.
    \item Finally, for $n_\Lambda=78$ (or $n_\Lambda$ close to it), we get charge $-e$ at the north pole and $+e$ at the South Pole. For $n_\Lambda>78$, i.e., $L>117$, the unprojected CFE is annihilated upon LLL projection.   
\end{enumerate}

Note that the unprojected state in panel (b) always consists of a charge $-e/3$ quasiparticle and a charge $+e/3$ quasihole. We also note that the QH at the South Pole is already in the LLL, so it is not affected by the LLL projection; the QH is thus simply a spectator, and all of the above properties belong to the CF in a high $\Lambda$L. Fig.~\ref{fig:cf-decay-schematic} is a schematic depiction of how a CF in a high $\Lambda$L splits into several fractionally charged QPs and QHs and even a whole electron.

Figure~\ref{fig:density-vs-lambda}(d) shows the excitation energy of the direct-projected CF exciton as a function of $n_\Lambda$ for $N=10$, 20, and 40. The energy rises with $n_\Lambda$; at the maximum index $n_\Lambda=2(N-1)$ the state coincides with the maximum-angular-momentum GMP mode, whose energy in the thermodynamic limit equals that of a far-separated electron-hole pair. We note that this is also the energy gap seen in STM~\cite{Peterson05,Gattu23,Pu23A,Hu23} and tunneling~\cite{Zhang17, Eisenstein92B} studies.

Corresponding density profiles, enclosed-charge distributions, and excitation energies at $\nu=1/5$ are presented in the Supplemental Material~\cite{SM}.

{\bf Direct vs. Jain-Kamilla projections:}
The belief so far has been that the direct and JK projections produce almost identical LLL wave functions. This is supported by the studies of ground and low energy excitations, where both projections produce wave functions that have very high overlap with the exact Coulomb eigenstates. Ref.~\cite{Balram24} showed numerically, however, that the relation between the GMP and single CF exciton modes was valid only for direct-projected CF excitons. The present work brings out qualitative differences between the two projection methods for high energy excitations, which we now discuss.

Let us consider the disk geometry for simplicity, where the state with angular momentum $m=-n$ in the $n^{\rm th}$ LL is given by $\bar{z}^n e^{-|z|^2/4}$, with the magnetic length set to unity, and $z_j=x_j-iy_j$. As discussed in detail in the literature~\cite{Jain97,Jain97b,Jain07}, the direct-projected CFE contains terms of the form
\begin{equation}
   \begin{aligned}
   &{\cal P}^{\rm Direct}_{\rm LLL} \bar{z_l}^n \prod_{j<k=1}^N(z_j-z_k)^2 e^{-\sum_p|z_p|^2/4}\\
   &\quad\propto e^{-\sum_p|z_p|^2/4}\left( -2{\partial\over \partial z_l}\right)^n \prod_{j<k=1}^N(z_j-z_k)^2\;.
   \end{aligned}
\end{equation}
Because the largest power of $z_l$ in $\prod_{j<k=1}^N(z_j-z_k)^2$ is $z_l^{2(N-1)}$, the LLL projection vanishes for $n>2(N-1)$.
In the JK projection, on the other hand, we encounter terms of the form
\begin{equation}
   \begin{aligned}
   &{\cal P}^{\rm JK}_{\rm LLL} \bar{z_l}^n \prod_{k\neq l}(z_l-z_k) e^{-|z_l|^2/4}\\
   &\quad\propto e^{-|z_l|^2/4}\left( -2{\partial\over \partial z_l}\right)^n \prod_{k\neq l}(z_l-z_k) \;.
   \end{aligned}
\end{equation}
The JK-projected CFE identically vanishes for $n>N-1$. These results remain valid for the spherical geometry as well.

Taking the example of a system of $N=40$ particles considered in the previous section, the direct-projected CF exciton exists for $n_{\Lambda}\leq 78$ whereas the JK-projected CF exciton for $n_{\Lambda}\leq 39$. This demonstrates that the JK and direct projections are qualitatively different for high energy CFEs. In particular, in contrast to the direct projection, the JK projection is unable to capture the GMP excitation consisting of a far separated electron-hole pair. This also demonstrates why the equivalence of the GMP mode to single CF excitons cannot be valid for JK-projected CF excitons. 

{\bf Discussion:}
This work was motivated by two seemingly irreconcilable results: (i) the GMP density wave mode can be expressed exactly as a linear combination of single CF excitons, and (ii) at large wave vectors the GMP mode produces a far separated pair consisting of a charge $-e$ electron and a charge $+e$ hole, whereas one expects a CF exciton to produce a far separated pair consisting of a CF quasiparticle of charge $-e/(2pn+1)$ and a CF quasihole of charge $+e/(2pn+1)$. By investigating the CF exciton at $\nu=1/3$, we find that a highly excited CF splits, upon projection into the LLL, into three charge $-e/3$ quasiparticles and two charge $+e/3$ quasiholes (Fig.~\ref{fig:cf-decay-schematic}); the former combine to produce a charge $-e$ electron while the latter combine with the quasihole to produce a charge $+e$ hole. That resolves the paradox at $\nu=1/3$. We have studied only the CF excitons of the $1/3$ and $1/5$ states, for which direct projection can be performed relatively straightforwardly, but we expect that a similar decay of a highly excited CF quasiparticle will occur for arbitrary Jain fractions. 

\begin{acknowledgments} 
We are grateful to Ajit Balram and G. J. Sreejith for insightful discussions. M.G. was supported in part by the National Science Foundation under Grant No.~DMR-2404619 and by a 2025/2026 Rising Researcher Grant from the Penn State Institute for Computational and Data Sciences (RRID:~SCR\_025154).  The authors of this work also recognize the Institute for providing access to computational research infrastructure within the Roar Core Facility (RRID:~SCR\_026424). M.G. thanks the Lodha Theoretical Physics Institute for their hospitality during the summer of 2026. Anthropic's Claude (Fable and Opus) assisted with the preparation of figures and with the coordination and execution of the numerical calculations.
\end{acknowledgments}

    \onecolumngrid
  \clearpage
  \begin{center}
    {\large\bfseries Supplemental Material for ``Spontaneous Decay of High Energy Composite Fermions''\par}
    \vspace{10pt}
    Mytraya Gattu and J. K. Jain
  \end{center}
  \vspace{10pt}
\setcounter{figure}{0}
\renewcommand{\thefigure}{S\arabic{figure}}
\renewcommand{\theHfigure}{S\arabic{figure}}
We present the density profiles, enclosed electric charge, and Coulomb excitation energies of maximum-weight CF excitons at filling factor $\nu=1/5$. For $N$ electrons at flux $2Q=5(N-1)$, these states have $L=L_z=N-1+n_\Lambda$ and survive direct LLL projection for $1\leq n_\Lambda\leq4(N-1)$. Figure~\ref{fig:sm-nu15} compares direct-projected and unprojected density and charge profiles and shows the projected excitation energies.

\begin{figure}[h!]
  \centering
  {\large $\nu=1/5$}\par\vspace{2pt}
  \begin{tabular}{@{}l@{\hspace{\columnsep}}l@{}}
    \textbf{(a)} & \textbf{(b)} \\
    \includegraphics[width=246pt]{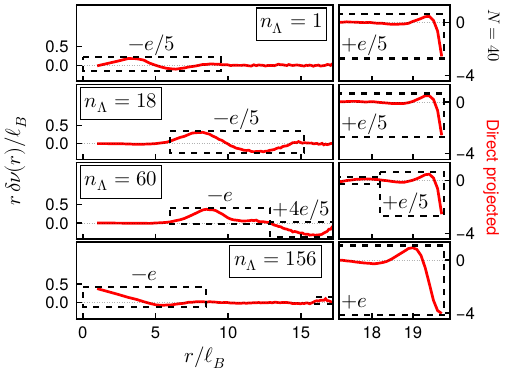} &
    \includegraphics[width=246pt]{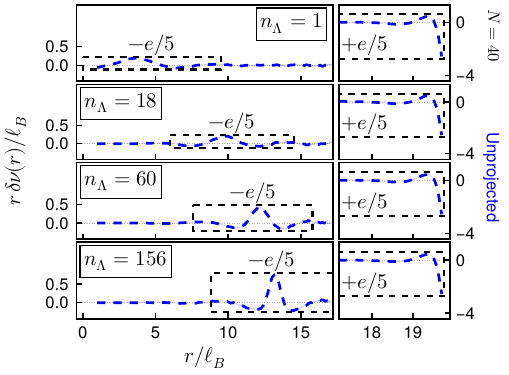} \\[4pt]
    \textbf{(c)} & \textbf{(d)} \\
    \includegraphics[width=246pt]{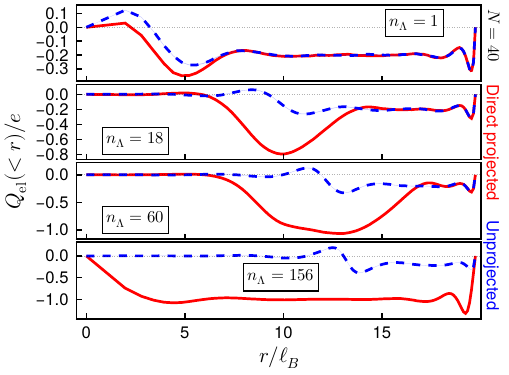} &
    \includegraphics[width=246pt]{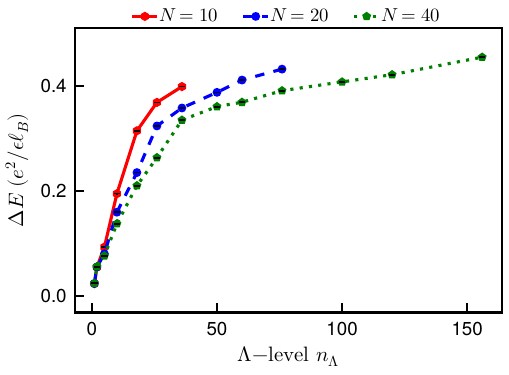}
  \end{tabular}
  \caption{Same as Fig.~\ref{fig:projection-density-charge} of the main text, but at $\nu=1/5$, with $n_\Lambda=1,18,60,156$ in (a), (b) and (c).}
  \label{fig:sm-nu15}
\end{figure}
\clearpage

\end{document}